\documentstyle [twocolumn,aps]{revtex}%

\begin{document}
\draft

\title{On the isospin dependence of the mean spin-orbit field in nuclei}

\author{V.I.\thinspace Isakov$^1$, K.I.\thinspace Erokhina$^2$,
H.\thinspace Mach$^3$, M.\thinspace Sanchez-Vega$^3$,
B.\thinspace Fogelberg$^3$}

\address{$^1$ Petersburg Nuclear Physics Institute,
Russian Academy of Sciences, Gatchina 188300, Russia}

\address{$^2$ Physicotechnical Institute,
Russian Academy of Sciences, St.Petersburg 194021, Russia}

\address{$^3$ Department of Radiation Sciences, Uppsala University,
 Nyk\"{o}ping S-61162, Sweden}


\maketitle
\vspace{-0.4cm}%
{\it%
By the use of the latest experimental data on the spectra of $^{133}$Sb
and $^{131}$Sn and on the analysis of properties of other odd nuclei
adjacent to doubly magic closed shells the isospin dependence of a mean
spin-orbit potential is defined. Such a dependence received the explanation
in the framework of different theoretical approaches.

}%
\vspace{0.4cm}%

\narrowtext

Recent experimental results \cite{1,2,3} on
nuclei close to $^{132}$Sn have lead to the determination of
a nearly complete set of neutron and proton single-particle
orbitals and to the establishment of some of their important
statical and dynamical properties. In particular, the new
results \cite{2} for $^{133}$Sb include information on
energies of proton single particle states above the $Z=50, N=
82$ shells as well as important knowledge about the decay properties
of these states.
In view of new data on single particle levels at $^{132}$Sn
we performed an analysis \cite{isa2000}
of the available information
on such states in strongly magical nuclides,
with special attention to the magnitudes of the spin-orbit splittings
and their isospin dependence. This question is important since such a
dependence could be one of the factors contributing to significant
structural changes in nuclides having an extreme neutron excess.

In \cite{2} the energy of the $3/2^+$ level in $^{133}$Sb was
measured to be 2.44 MeV. By using this value and also
the previous data on the spectrum of
single particle excitations in nuclei close to
$^{132}$Sn (see \cite{1} and \cite{15,16,17,18}) the values of
spin-orbit splittings of the $2d$ levels both in proton and neutron
systems of $^{132}$Sn were defined. The splitting was found to
be 1.48 MeV for protons and 1.65 MeV for neutrons, i.e. the
neutron spin-orbit splitting is somewhat larger for neutrons than
for protons. At the same time it was noted in \cite{2} that for
nuclei close to $^{208}$Pb the situation is the opposite, in any
case for the first glance. So, from the spectra of single
particle levels in $^{209}$Bi and $^{207}$Pb it follows that the
spin-orbit splitting of the proton $2f$ orbit is equal to 1.93
MeV, while for neutrons it is 1.77 MeV. However it follows  from
the experiment that the neutron $2f7/2$ state in $^{207}$Pb is
strongly fragmented. So, the conclusions of the work \cite{2}
refer only to the lowest, though the strongest component of this
state. Identifying in the spirit of \cite{Cohen63},\cite{Bar70}
the true single particle energy
of the $2f7/2$ state with the weighted average of $7/2^-$ energy
levels, the weight being the spectroscopic factors of the (d,t)
reaction on $^{208}$Pb \cite{19},
we obtain the real excitation energy of this state equal to 2.70
MeV (instead of 2.34 MeV). This corresponds to the value of
neutron spin-orbit splitting of the $2f$ orbit equal to 2.13 MeV,
i.e. as for the $2d$ orbit in $^{132}$Sn a little larger than that for
protons. The above statement is fortified by the analysis of the
$3p$ spin-orbit splitting near $^{208}$Pb. Such a splitting is
equal to 0.85 MeV for protons (taking into account the
fragmentation of $3p1/2$ level) while for neutrons it is 0.90 MeV,
i.e. a little larger than for protons.

The systematics of single particle energies in $^{208}$Pb and
$^{132}$Sn available by now is presented in Tables 1 and 2. In
composing these tables the energies of the particle and hole
states closest to the Fermi-level were determined from the
differences of binding energies of the core and the
corresponding adjacent odd nuclei:
$\varepsilon$(particle)=B(core)--B(core+nucleon) and
$\varepsilon$(hole)=B(core--nucleon)--B(core), using the
experimental data from \cite{20}. The energies of orbitals more
remote from the Fermi-level were defined after that by the
addition (subtraction) of the excitation energies \cite{1},
\cite{2}, \cite{15}--\cite{18}, \cite{21}, \cite{22} of the
corresponding orbitals in the adjacent odd nuclei, accounting
for the fragmentation of states, if the corresponding data are
available. It really follows from Tables 1 and 2 that the
neutron spin-orbit splitting in magical nuclei $^{208}$Pb and
$^{132}$Sn, both having considerable neutron excess as compared
to the number of protons, is larger than the corresponding proton
splitting by about  $\sim$ 10\%.

One can see a completely different picture in the $N=Z$ nuclei; see
Tables $3 \div 5$. Here the spin-orbit splittings of the $1d$ and
$1p$ levels in $^{16}$O \cite{28}, \cite{29} as well as those of $1f$
and $2p$ levels in $^{40}$Ca \cite{30} are practically equal,
another time suggesting the concept of isobaric invariance in nuclei.
By the present time the experimental data on the structure of
single particle states in $^{100}$Sn are absent. However the
work \cite{31} presents their spectrum obtained from the
extrapolation of the properties of nuclei with less neutron
deficiency (see Table 5). A conclusion based on these data is
that, within the errors the spin-orbit splitting of
the $1g$ orbit is also equal for protons and neutrons.

Turning to the theoretical interpretation \cite{isa2000} of data on the
spin-orbit splitting we shall first recall that from the
point of view of multiparticle theory the
average spin-orbit potential has it's origin in the
pair spin-orbit interaction between nucleons (with the
tensor forces also giving some contribution in the second
order perturbation theory). On the level of qualitative
arguments it was noted by ref. \cite{32} that due
to the symmetry properties one should expect the neutron
spin-orbit splitting somewhat larger than that for protons.
However, at that time the absence of experimental data did
not permit to make a meaningful comparison with
measurements. With the presently available
data including those obtained by us in \cite{2} we can
fill this gap, giving also some quantitative
considerations.

The two-body spin-orbit interaction differs from zero only in the
states with total spin $S=1$. Neutron-neutron and proton-proton
systems have the total isospin $T=1$ and thus due to the Pauli
principle have odd values of the relative orbital momentum $L$
(really, $L=1$). At the same time, the neutron-proton system is
composed from the $T=0$ and $T=1$ states with equal weights,
correspondingly having $L=0$ and $L=1$. However, the spin-orbit
interaction is absent in states with $L=0$. As a result, the pair
spin-orbit $np$ interaction is half as strong as that in $pp$ or
$nn$-systems.

If $U_{\ell s}(n)$ and $U_{\ell s}(p)$ are the values presenting
the magnitudes of the mean spin-orbit field for neutrons and
protons and $\vartheta(T=1, S=1, L=1)$
is a quantity representing the parameter of the
pair spin-orbit interaction in a state with $T=1, S=1, L=1$ then
the above discourse leads to $$U_{\ell s}(n)\sim \vartheta(1, 1, 1)
\left(N+\frac 12 Z\right) \equiv \vartheta \cdot\left(A-\frac
Z2\right) \, {\rm and}$$ $$ U_{\ell s}(p)\sim \vartheta(1, 1, 1)
\left(\frac N2+Z\right) \equiv \vartheta \cdot\left(A-\frac N2\right)
.\eqno{(1)}$$

As the spin-orbit splitting $\Delta_{\ell s}^{(n,p)}\sim
U_{\ell s}(n,p)$, the relative difference "$\varepsilon$" of the neutron
and proton spin-orbit splittings is given by the expression:

$$\varepsilon=\frac{\Delta_{\ell s}^{(n)}-\Delta_{\ell s}^{(p)}}
{(\Delta_{\ell s}^{(n)}+\Delta_{\ell s}^{(p)})/2}
=\frac 23\,
\frac{N-Z}{A}\,.\eqno{(2)}$$

On the other side, if we express the parameter of the
spin-orbit mean field in the form

$$U_{\ell s}(\tau_{3})=V_{\ell s}\left(1+\frac{1}{2}\,
\beta_{\ell s}\frac{N-Z}{A}\cdot
\tau_{3}\right)\,,\eqno{(3)}$$
where $\tau_{3}=-1$ for neutrons and $\tau_{3}=+1$ for protons, then

$$\varepsilon=-\beta_{\ell s}\frac{N-Z}{A}\,,\eqno{(4)}$$
i.e. it follows from the comparison of (2) and (4) that
$\beta_{\ell s}=-2/3$.

One can also make the evaluation of the isotopical dependence of
spin-orbit interaction in the Hartree approximation starting from the
Dirac phenomenology with meson-nucleon interactions \cite{Wal74}.
There one obtains (see for example \cite{Brock78}--\cite{Yosh99} and
the references therein) the Skyrme-type single particle equation for
a nucleon having the effective mass $m_{N}^{*}$.
Here we concentrate on the isotopical dependence of the spin-orbit
potential having the form, see for example \cite{Rein86}--\cite{Bir98}:

$$\hat{U}_{\ell s}= \frac{\lambda_{N}^{2}}{2}\, \frac{1}{r}\,
\{(\frac{m_{N}}{m_{N}^{*}})^2 \frac{d}{dr}\lbrack ( V_{\omega}^{0} -
S_{\sigma,{\sigma}_0}^{0})-\,$$
$$- (V_{\rho}^{1}-S_{\delta,\sigma,{\sigma}_0}^{1})\cdot
\tau_{3} \rbrack -2k(\frac{m_{N}}{m_{N}^{*}})
\frac{d}{dr} V_{\rho}^{1} \cdot \tau_{3} \}\,
\hat{\bf \ell} \cdot \hat{\bf s}. \, \eqno{(5)}$$

Here $V = V^{0}-\tau_{3}\cdot V^{1}$ \,and \,
$S = S^{0} -\tau_{3}\cdot S^{1}$ are the vector and
scalar fields due to corresponding mesons,
$m_{N}^{*} = m_{N} + \frac{1}{2}(S-V)$, while "$k$" is the
ratio of tensor to vector coupling constants of $\rho$-meson.
In \cite{Bir98} the meson-nucleon coupling constants defining
the $V$ and $S$ fields were borrowed from the Bonn
$NN$ boson exchange potential \cite{Mac87}, where $\sigma$ and
${\sigma}_0$ are scalar mesons imitating the 2$\pi$ exchange
in the $NN$- systems with $T$=1 and $T$=0 correspondingly.
At the same time, in some other works  (see for example
\cite{Rein86}--\cite{Koepf91}) the mentioned constants were
defined from the description of global nuclear properties,
with inclusion of the $\sigma^{3}$ and $\sigma^{4}$ terms in
the Lagrangian density (one $\sigma$-meson with the same
characteristics for $T$=1 and $T$=0 channels was used which
leads to zero contribution of this meson to  $S^1$
in formula (5), the tensor term was not included in the
$\rho$-meson vertex in the cited works). Taking into account
that the radial dependence of the $(m_{N}/m_{N}^{*})$ is much
weaker than that of $V$ and $S$, which are considered as
proportional to the density having the Fermi--function form,
one can approximately present formula (5) as

$$\frac{1}{x}\, \frac{df}{dx} \cdot V_{\ell s} \left( 1+
\frac{1}{2}\,\beta_{\ell s} \frac{N-Z}{A} \cdot\tau_{3} \right)
\hat{\bf \ell} \cdot \hat{\bf s} \,;$$
$$f =\lbrack 1+exp\,(\frac{x-R}{a}) \rbrack^{-1} \,. \eqno{(6)}$$

Calculating the $V$ and $S$ magnitudes in the
center of nuclei at the values of vector and scalar densities
${\rho}_{v}$ = 0.17, ${\rho}_{s}$ = 0.16,\, ${\rho}_{v}^{-}$ =
0.17\,$(N-Z)/A$, ${\rho}_{s}^{-}$ = 0.16\,
$(N-Z)/A$ (all in {fm}$^{-3}$) , using the coupling parameters
from \cite{Bir98}, \cite{Mac87} and taking into
account the isotopic dependence of $m_{N}/m_{N}^{*}$, we obtain
$V_{\ell s}$ = 33.6 MeV and \, $\beta_{\ell s}$ = -- 0.40 with "$x$"
in the units of fm. If we use the set of parameters NL2 from
\cite{Lee86,Koepf91} than we have $V_{\ell s}$ = 31.3 MeV,
$\beta_{\ell s}$ = -- 0.43. At the same time, the set NL1 from
\cite{Rein86,Koepf91} giving small values of effective masses
leads to $V_{\ell s}$ $\approx$ 50 MeV and
$\beta_{\ell s}$ $\approx$ -- 1.3. As the $V^{1}, S^{1}$
magnitudes are proportional to ${\rho}_{v}^{-},{\rho}_{s}^{-}$
both the formulas (5) and (6) give the spin-orbit splitting
equal for protons and neutrons in the $N=Z$ nuclei.
It should be mentioned that
in all cases the value of $\beta_{\ell s}$ is always
negative and is defined mainly or entirely by contribution
of a $\rho$-meson.

We note here that the data on spin-orbit splittings of the
$2d$ states in $^{132}$Sn as well as on the splittings of $2f$ and $3p$
levels in $^{208}$Pb lead to effective values of $\beta_{\ell s}$
equal to $-0.55$, $-0.60$ and $-0.27$ correspondingly.

Let single particle levels be generated by the potential
$$\hat{U}(x,\hat \sigma,\tau_{3})=U_{0}(\tau_{3})f(x)\,+$$
$$+\, \frac{U_{\ell s}(\tau_{3})}{x}\,\frac{df}{dx}\,
\hat{\bf \ell} \cdot \hat{\bf s} +
\frac{(1+\tau_{3})}{2}\,U_{Coul}\,, \eqno{(7)}$$

\noindent
where $U_{0}(\tau_{3})=
V_{0}(1+\frac{1}{2}\beta\frac{N-Z}{A}\cdot \tau_{3})$;
$U_{\ell s}$ and $f(x,a,R)$ are defined by eq. (3) and (6),
$R=r_{0}A^{1/3}$, while $U_{Coul}(x,R_{c},Z)$ presents the
potential of the uniformly charged sphere with the charge $Z$
and radii $R_{c}=r_{c}A^{1/3}$.

In works \cite{9} --\cite{13} single particle levels were described
by using the set of parameters $V_{0}=-51.5$ MeV, $r_{0}$ =
1.27 fm, $V_{\ell s}$ =33.2 MeV, $a(p)$= 0.67 fm, $a(n)$ = 0.55 fm
and $\beta_{\ell s}$ =$\beta$ =1.39, which on the average described
the spectra of single particle states in nuclei from $^{16}$O
to $^{208}$Pb. This set of parameters we call as the "Standard"
one. With the appearance of new experimental data on the
single-particle levels we performed a new determination of parameters
entering formula (7). They were defined by using the Nelder--Mead
method \cite{Nel67} through minimization of the root-mean square
deviation
$$\delta=\sqrt{\frac 1N\sum\limits_{k=1}\limits^{N}
(\varepsilon_k^{\rm theor}-\varepsilon_k^{\rm exp})^2}\,.
\eqno{(8)}$$

The computation demonstrated a very small sensitivity of results
to the values of $r_c$, which was adopted by us the same as
before, $r_c=1.25$ fm. The minimization of $\delta$ held for
all nuclei presented in Tables $1\div 5$ with $r_c=1.25$ fm and
different values of $r_0$ showed that the minimum in all cases
corresponds to $r_0\approx 1.27$ fm which also coincides with
the value adopted by us before. The values $r_c=1.25$ fm and
$r_0=1.27$ fm were fixed in further calculations.

As was noted above, the optimal relation of proton to neutron
spin-orbit splitting corresponds to $\beta_{\ell s}\sim-0.6$.
The fourth column of Tables 1 and 2 (variant 1 of calculations)
presents the values of theoretical energy levels obtained
in the optimization with fixed values of $\beta_{\ell s}=-0.6$,
$a_p=0.67$ fm and $a_n=0.55$ fm.

The fifth column of Tables $1\div 2$ (variant 2) presents the
results of optimization with two fixed parameters: $a_p=0.67$ fm
and $a_n=0.55$ fm.

Variant 3 corresponds to optimization at fixed $\beta_{\ell
s}=-0.6$, while variant 4 presents the results with no parameters
fixed.

We see that the optimal values of $V_0$, $V_{\ell s}$ and
$\beta$ (see formula (7)) are very close to the "Standard"
ones, with small dispersion from nuclei to nuclei. The
magnitudes of the diffusinesses "$a$" vary more strongly,
differing by about 10 $\div$ 15$\%$ from their "standard" values.
As one can see from the comparisons of the "Stnd" with "Set 1"
and "Set 3" with "Set 4" fittings, the contribution of
$\beta_{\ell s}$ that defines the isospin dependence
of spin-orbit splitting to $\delta$ is small. It is more
reasonable to define it's value not from minimization of
$\delta$, but from experimental and theoretical
arguments mentioned above.
This conclusion is confirmed by the result of \cite{Koura20},
where different fittings gave diverse (in magnitudes and signs)
values of parameter defining the linear in $(N-Z)/A$ contribution
to the spin-orbit term (energies with maximal values of
spectroscopic factors were used as the input ones in these
fittings).

The energies of levels in nuclei with $N=Z$ (see Tables 3,4,5)
are independent on $\beta$ and $\beta_{\ell s}$. Here the optimization
was performed twice, once with fixed values of $a_n=0.55$ fm
and $a_p=0.67$ fm with subsequent definition of $V$ and $V_{\ell s}$
(variant 1) and once without fixing some parameters (variant 3).

The results of calculations presented in Tables 1 to 5 refer
also to some levels having positive energies, i.e. to unbound, but
sub-barrier states. In these cases we present here only the real
part of single particle energies having here really very small
decay widths.

For comparison of the results obtained by using the
empirical potential (7) with those using the microscopical
procedure we also performed
Hartree-Fock calculations with the SIII
interaction (last two columns of Tables 1 to 5). The results
of our self-consistent calculations were obtained by considering
the contribution of a single-particle part of the center-of-mass
energy and taking into account the Coulomb exchange term in the
Slater approximation. The SIII(1) data correspond to calculations
taking into account all the terms of the energy functional contributing
to spin-orbit splitting, while the SIII(2) results were obtained by
omitting the spin density terms in the spin-orbit potential. In
the last case our results are close to that from the work \cite{37}
for $^{208}$Pb, $^{132}$Sn and $^{100}$Sn nuclei. We  see that the
results obtained in the framework of the Hartree-Fock method also
demonstrate that the neutron spin-orbit splittings of the  $2d$
orbit in $^{132}$Sn as well as of the
$2f$ and $3p$ orbits in $^{208}$Pb are larger than those for protons
and correspond to effective $\beta_{\ell s}$ in the interval of $-0.9 \div
-0.6$. 
One can mention
here that the difference between the neutron and proton spin-orbit
splittings is reproduced by using the simple parameterization of
Skyrme forces. More careful parameterization enables to reproduce
\cite{Sharma95} the anomalous kink of isotopic shifts in Pb isotopes.

Tables 3 and 4 presented below refer to $^{16}$O and $^{40}$Ca, which
are the spin saturated nuclei. In these cases the spin density terms
in practice do not contribute the spin-orbit splitting (the
corresponding contributions in these cases are due only to small
distinction of radial wave functions of spin-orbit partners). One can
see, that the SIII-1 and SIII-2 calculations give here almost
similar results.

We conclude that our analysis,
based both on experimental data and on
different theoretical approaches has defined the isotopic dependence
of the nuclear mean field spin-orbit splitting. The splitting becomes
larger for neutrons than for protons in nuclei with $N>Z$. Importantly,
the theoretical analysis shows that the difference between the neutron
and proton splittings becomes saturated which precludes very large
differences. The rather modest difference seen in the $^{132}$Sn
region is already about $25\%$ of the saturation value, showing that the
isospin dependence can not introduce major structural changes even
in extreme cases of neutron excess.

This work was supported by
the Swedish Natural Research Council,
the Royal Swedish Academy of Sciences
and the Russian Foundation of Fundamental Research (grant No.
96-15-96764).  The authors are grateful to B.L. Birbrair for
discussion.

{\bf Table 1. Single particle states of $^{208}$Pb.}
\begin{center}
{\footnotesize\begin{tabular}{|c|c|c|c|c|c|c|c|c|} \hline \hline
$n\ell j$&$\varepsilon_{exp}$&Stnd&Set 1&Set 2&Set 3&Set 4&
SIII-1 & SIII-2\\  \hline\hline
$n 3d_{3/2}$&-1.40&-0.32&-0.02&-0.23&-0.96&-0.99&0.38&0.42\\
    \hline
$n2g_{7/2}$&-1.44&-0.79&-0.18&-0.65&-0.89&-1.14&0.01&0.14\\ \hline
$n4s_{1/2}$&-1.90&-0.80&-0.70&-0.74&-1.63&-1.51&-0.08&-0.06\\
\hline
$n1j_{15/2}$&-$2.09^*$&-2.42&-3.05&-2.31&-2.23&-1.55&-1.41&-1.93\\
\hline
$n3d_{5/2}$&-2.37&-1.50&-1.45&-1.40&-2.35&-2.13&-0.39&-0.38\\
\hline
$n1i_{11/2}$&-3.16&-4.24&-3.37&-4.05&-2.71&-3.33&-3.37&-2.77\\
\hline
$n2g_{9/2}$&-3.94&-3.71&-3.82&-3.59&-4.24&-3.88&-2.91&-2.97\\
\hline
$n3p_{1/2}$&-7.37&-7.32&-6.94&-7.17&-7.59&-7.61&-7.21&-7.13\\ \hline
$n2f_{5/2}$&-7.94&-8.42&-7.87&-8.25&-8.17&-8.38&-8.59&-8.44\\ \hline
$n3p_{3/2}$&-8.27&-8.18&-8.03&-8.04&-8.59&-8.43&-8.18&-8.15\\ \hline
$n1i_{13/2}$&-9.00&-9.21&-9.62&-9.08&-8.84&-8.31&-9.73&-10.21\\
\hline
$n2f_{7/2}$&-$10.07^*$&-10.57&-10.57&-10.43&-10.72&-10.46&-11.21
& -11.24\\ \hline
$n1h_{9/2}$&-10.78&-12.06&-11.35&-11.87&-10.60&-11.09&-13.16&-12.67\\
\hline
\hline
$p3p_{1/2}$&$0.17^*$&0.63&0.43&0.72&0.29&0.47&2.79&2.88\\ \hline
$p3p_{3/2}$&-0.68&-0.45&-0.46&-0.35&-0.58&-0.69&1.99&2.03\\ \hline
$p2f_{5/2}$&-0.97&-0.68&-1.03&-0.60&-1.03&-0.61&0.60&0.74\\ \hline
$p1i_{13/2}$&-2.19&-2.86&-2.37&-2.71&-1.94&-2.78&-1.20&-1.53\\
\hline
$p2f_{7/2}$&-2.90&-3.38&-3.24&-3.26&-3.21&-3.53&-1.64&-1.66\\ \hline
$p1h_{9/2}$&-3.80&-4.60&-5.11&-4.53&-4.71&-4.01&-4.68&-4.24\\ \hline
$p3s_{1/2}$&-8.01&-7.76&-7.86&-7.67&-7.87&-7.87&-7.39&-7.33\\ \hline
$p2d_{3/2}$&-8.36&-8.41&-8.66&-8.32&-8.59&-8.30&-8.64&-8.51\\ \hline
$p1h_{11/2}$&-9.36&-9.33&-8.99&-9.18&-8.60&-9.21&-9.35&-9.65\\ \hline
$p2d_{5/2}$&-10.04$^*$&-10.10&-10.05&-9.98&-9.96&-10.15&-10.29
&-10.28\\ \hline
$p1g_{7/2}$&-12.18$^*$&-12.07&-12.45&-11.99&-12.08&-11.58&-13.94&-13.59\\
\hline\hline
\end{tabular}}
\end{center}

{\footnotesize
The "standard" set of parameters corresponds to $V_0=-51.50$
MeV, $V_{\ell s}=33.2$ MeV, $\beta=\beta_{\ell s}=+1.39$,
$a_p=0.67$ fm, $a_n=0.55$ fm and $\delta=0.604$ MeV.

Set "1" corresponds to $V_0=-51.39$ MeV, $V_{\ell s}=33.1$ MeV,
$\beta=1.43$ with $\beta_{\ell s}=-0.6$, $a_p=0.67$ fm,
$a_n=0.55$ fm fixed; $\delta=0.654$ MeV.

Set "2" corresponds to $V_0=-51.34$ MeV, $V_{\ell s}=33.1$ MeV,
$\beta=1.40$, $\beta_{\ell s}=1.26$ with $a_p=0.67$ fm,
$a_n=0.55$ fm fixed; $\delta=0.593$ MeV.

Set "3" corresponds to $V_0=-51.99$ MeV, $V_{\ell s}=32.7$ MeV,
$\beta=1.36$, $a_p=0.73$ fm,
$a_n=0.72$ fm with $\delta=0.369$ MeV;
$\beta_{\ell s}=-0.6$ is fixed.

Set "4" corresponds to $V_0=-51.93$ MeV, $V_{\ell s}=35.2$ MeV,
$\beta=1.38$, $\beta_{\ell s}=1.76$,  $a_p=0.73$ fm,
$a_n=0.72$ fm;  $\delta=0.366$ MeV.

Here and below experimental single particle energies marked by
asterisks were obtained using the averaging over the
spectroscopic factors.}

\newpage

{\bf Table 2. Single particle states of $^{132}$Sn.}
\begin{center}

{\footnotesize\begin{tabular}{|c|c|c|c|c|c|c|c|c|} \hline \hline
$n\ell j$&$\varepsilon_{exp}$&Stnd&Set 1&Set 2&Set 3&Set 4&
SIII-1& SIII-2 \\ \hline\hline
$n2f5/2$&-0.58&0.36&0.73&0.46&0.22&-0.01&0.67&0.79\\ \hline
$n3p1/2$&(-0.92)&-0.13&-0.48&-0.09&-0.55&-0.61&0.16&0.20\\
\hline
$n1h9/2$&-1.02&-1.61&-0.84&-1.38&-0.47&-0.97&-0.72&-0.02\\
\hline
$n3p3/2$&-1.73&-0.78&-0.88&-0.77&-1.42&-1.32&-0.16&-0.14\\
\hline
$n2f7/2$&-2.58&-2.18&-2.55&-2.21&-2.84&-2.52&-1.67&-1.71\\
\hline
$n2d3/2$&-7.31&-7.74&-7.45&-7.62&-7.63&-7.77&-8.42&-8.26\\ \hline
$n1h11/2$&-7.55&-7.11&-7.96&-7.23&-7.33&-6.60&-7.69&-8.23\\
\hline
$n3s1/2$&-7.64&-7.68&-7.73&-7.64&-8.03&-7.93&-8.26&-8.21\\ \hline
$n2d5/2$&-8.96&-9.66&-9.94&-9.66&-9.98&-9.69&-10.71&-10.71\\
\hline
$n1g7/2$&-9.74&-10.56&-10.04&-10.39&-9.51&-9.81&-11.92&-11.32\\
\hline
\hline
$p3s1/2$&(-6.83)&-6.84&-6.87&-6.80&-6.64&-6.70&-4.97&-4.90\\
\hline p1h11/2&-6.84&-7.32&-6.66&-7.46&-6.77&-7.48&-5.64&-6.01\\
\hline
$p2d3/2$&-7.19&-6.86&-7.20&-6.74&-7.07&-6.72&-5.93&-5.77\\
\hline
$p2d5/2$&-8.67&-9.36&-9.20&-9.37&-9.04&-9.30&-7.88& -7.88\\
\hline
$p1g7/2$&-9.63&-9.84&-10.41&-9.66&-10.60&-9.81&-10.08 &-9.56\\
\hline
$p1g9/2$&-15.71&-14.91&-14.46&-15.00&-14.57&-15.02&-15.03&-15.36\\
\hline
$p2p1/2$&-16.07&-16.01&-16.22&-15.92&-16.14&-15.91&-16.68&-16.55\\
\hline \hline
\end{tabular}}
\end{center}

{\footnotesize
Stnd:  $\delta=0.589$ MeV.

Set 1: $V_0=-51.56$ MeV, $V_{\ell s}=33.3$ MeV,
$\beta=1.39$,  $\delta=0.638$ MeV.

Set 2: $V_0=-51.44$ MeV, $V_{\ell s}=34.8$ MeV,
$\beta=1.39$, $\beta_{\ell s}=1.35$, $\delta=0.575$ MeV.

Set 3:  $V_0=-51.55$ MeV, $V_{\ell s}=32.4$ MeV,
$\beta=1.31$, $a_p=0.63$ fm,
$a_n=0.66$ fm, $\delta=0.546$ MeV.

Set 4: $V_0=-51.56$ MeV, $V_{\ell s}=34.1$ MeV,
$\beta=1.34$, $\beta_{\ell s}=1.33$,  $a_p=0.65$ fm,
$a_n=0.66$ fm,  $\delta=0.478$ MeV.

Note that some theoretical works \cite{44} postulate that the neutron
1i13/2 state at $^{132}$Sn is only 1.9 MeV above the n2f7/2
level. Our calculations unequivocally demonstrate, that this state
lies considerably higher, with it's energy equal to +0.55,
+1.59 and +1.02 MeV correspondingly for the "Stnd", SIII-1
and SIII-2 variants.}

\vspace{2.0cm}
{\bf Table 3. Single particle levels of $^{16}$O.}
\begin{center}

\begin{tabular}{|c|c|c|c|c|c|c|}\hline\hline
$n\ell j$&$\varepsilon_{exp}$&Stnd&Set 1&Set 3&SIII-1 &SIII-2\\
\hline\hline
$n1d3/2$&(0.94)&0.89&0.18&0.20&0.66&0.67\\ \hline
$n2s1/2$&-3.27&-3.59&-3.89&-3.31&-2.88&-2.87\\ \hline
$n1d5/2$&-4.14&-6.97&-6.85&-6.41&-6.87 &-6.89\\ \hline
$n1p1/2$&-15.67&-15.06&-16.05&-16.33&-14.58 &-14.56\\ \hline
$n1p3/2$&(-21.84)&-19.98&-20.25&-20.10&-20.58 &-20.59\\ \hline
\hline
$p1d3/2$&(4.40)&3.76&2.92&3.48&3.55&3.56\\ \hline
$p2s1/2$&-0.11&-0.89&-1.14&0.22&0.03&0.03\\ \hline
$p1d5/2$&-0.60&-2.76&-2.67&-2.97&-3.57 &-3.59\\ \hline
$p1p1/2$&-12.13&-9.95&-10.87&-12.60&-11.17 &-11.15\\ \hline
$p1p3/2$&(-18.45)&-14.66&-14.90&-16.40&-17.07 &-17.08\\
\hline\hline
\end{tabular}
\end{center}
{\footnotesize
Set 1: $V_0=-52.21$ MeV, $V_{\ell s}=28.6$ MeV;
$a_p=0.67$ fm, $a_n=0.55$ fm  are fixed.

Set 3: $V_0=-51.40$ MeV, $V_{\ell s}=25.7$ MeV,
$a_p=0.45$ fm, $a_n=0.50$ fm.
}

\newpage

{\bf Table 4. Single particle states of $^{40}$Ca.}
\begin{center}

\begin{tabular}{|c|c|c|c|c|c|c|}\hline\hline
$n\ell j$&$\varepsilon_{exp}$&Stnd&Set 1&Set 3&SIII-1 &SIII-2\\
\hline \hline
$n1f5/2$&-3.48&-2.57&-3.91&-3.54&-1.49 &-1.48\\ \hline
$n2p1/2$&-4.42&-3.35&-4.08&-4.69&-2.20 &-2.23\\ \hline
$n2p3/2$&-6.42&-5.71&-6.08&-6.57&-4.09 &-4.05\\ \hline
$n1f7/2$&-8.36&-10.43&-10.44&-9.72&-9.92 &-9.94\\ \hline
$n1d3/2$&-15.64&-16.21&-17.40&-16.43&-15.53 &-15.54\\ \hline
$n2s1/2$&-18.11&-16.51&-17.17&-17.00&-15.94 &-15.92\\ \hline
$n1d5/2$&-$21.64^*$&-21.08&-21.44&-20.52&-21.90 &-21.90\\ \hline
\hline
$p1f5/2$&3.86&4.92&3.79&3.41&4.90&4.91\\ \hline
$p2p1/2$&2.64&2.62&2.11&2.07&3.66&3.64\\ \hline
$p2p3/2$&0.63&0.89&0.60&0.45&2.23&2.26 \\ \hline
$p1f7/2$&-1.09&-2.19&-2.18&-2.85&-3.04 &-3.06\\ \hline
$p1d3/2$&-8.33&-7.11&-8.25&-9.01&-8.52 &-8.53\\ \hline
$p2s1/2$&-10.85&-8.18&-8.78&-9.30&-8.77 &-8.75\\ \hline
$p1d5/2$&-$14.33^*$&-12.05&-12.36&-13.19&-14.74 &-14.75\\
\hline\hline
\end{tabular}
\end{center}

{\footnotesize
Set 1: $V_0=-52.39$ MeV, $V_{\ell s}=27.9$ MeV;
$a_p=0.67$ fm and $a_n=0.55$ fm  are fixed.

Set 3: $V_0=-52.95$ MeV, $V_{\ell s}=28.2$ MeV,
$a_p=0.63$ fm, $a_n=0.68$ fm.}

\vspace{2.0cm}

{\bf Table 5. Single particle states of $^{100}$Sn.}
\begin{center}
\begin{tabular}{|c|c|c|c|c|c|c|}\hline\hline
$n\ell j$&$\varepsilon_{sys}$&Stnd&Set 1&Set 3&SIII-1 &SIII-2\\
\hline \hline
$n1h11/2$ &-8.6(5) &-8.66 &-9.01 &-8.72 &-6.35 &-6.87\\ \hline
$n2d3/2$ &-9.2(5) &-8.90 &-9.24 &-8.70 &-7.84 &-7.66\\ \hline
$n3s1/2$ &-9.3(5) &-9.16 &-9.53 &-9.13 &-7.58 &-7.52\\ \hline
$n1g7/2$&-10.93(20)&-11.64&-12.02&-11.23&-10.33 &-9.63\\ \hline
$n2d5/2$&-11.13(20)&-11.62&-11.97&-11.59&-10.07 &-10.10\\ \hline
$n1g9/2$&-17.93(20)&-17.23&-17.61&-17.21&-16.54 &-17.00\\ \hline
$n2p1/2$&-18.38(20)&-19.14&-19.53&-18.93&-19.08 &-18.93\\ \hline
\hline
$p1g7/2$&3.90(15)&3.88&3.54&2.70&3.38 &4.04\\ \hline
$p2d5/2$&3.00(80)&2.74&2.45&2.64&3.70 &3.69\\ \hline
$p1g9/2$&-2.92(20)&-2.01&-2.36&-3.66&-2.74 &-3.16\\ \hline
$p2p1/2$&-3.53(20)&-3.48&-3.84&-3.94&-4.80 &-4.65\\ \hline
$p2p3/2$&-6.38&-4.95&-5.31&-5.55&-6.22 &-6.18\\ \hline
$p1f5/2$&-8.71&-5.54&-5.92&-7.60&-8.43 &-7.89 \\
\hline\hline
\end{tabular}
\end{center}

{\footnotesize
Set 1: $V_0=-51.97$ MeV, $V_{\ell s}=33.5$ MeV;
$a_p=0.67$ fm and $a_n=0.55$ fm  are fixed.

Set 3: $V_0=-51.40$ MeV, $V_{\ell s}=35.6$ MeV,
$a_p=0.52$ fm, $a_n=0.56$ fm.}


\vspace{-0.2cm}


\begin{thebibliography}{99}
\vspace{-1.5cm}

\bibitem{1}
P. Hoff et al.,\, Phys. Rev. Lett. {\bf 77}, 1020 (1996).

\bibitem{2}
M. Sanchez-Vega, B. Fogelberg, H. Mach, R.B.E. Taylor, A. Lindroth,
Phys. Rev. Lett. {\bf 80}, 5504 (1998).

\bibitem{3}
N. J. Stone et al., Phys. Rev. Lett. {\bf 78}, 820 (1997).

\bibitem{isa2000}
V. I. Isakov, K. I. Erokhina, H. Mach, M. Sanchez-\,Vega, B. Fogelberg,
Preprint PNPI, Gatchina, No 2375 (2000).

\bibitem{15}
T. Bjornstad et al., Nucl.Phys. {\bf A453}, 463 (1986).

\bibitem{16}
Yu.V. Sergeenkov, Yu.L. Khazov, Nucl.Data Sheets {\bf 72}, 487 (1994).

\bibitem{17}
Yu.V. Sergeenkov, V.M. Sigalov, Nucl.Data Sheets {\bf 49}, 639 (1986).

\bibitem{18}
G.A. Stone, S.H. Faller, W. Walters, Phys.Rev. C {\bf 39}, 1963 (1989).

\bibitem{Cohen63}
B.L. Cohen, R.H. Fulmer, A.L. McCarthy, P. Mukherjee, Rev.Mod.Phys
{\bf 35}, 332 (1963).

\bibitem{Bar70}
M. Baranger,\, Nucl.Phys. {\bf A149}, 225 (1970).

\bibitem{19}
R.A. Moyer, B.L. Cohen, R.C. Diehl,
Phys.Rev. C {\bf 2}, 1898 (1970).

\bibitem{20}
G. Audi, A.H. Wapstra,\, Nucl.Phys. {\bf A565}, 1 (1993).

\bibitem{21}
M.J. Martin,\, Nucl.Data Sheets {\bf 63}, 723 (1991).

\bibitem{22}
M.J. Martin,\, Nucl.Data Sheets {\bf 70}, 315 (1993).

\bibitem{28}
D.R. Tilley, H.R. Weller, C.M. Cheves,
Nucl.Phys. {\bf A564}, 1 (1993).

\bibitem{29}
F. Ajzenberg-Selove,\,
Nucl.Phys. {\bf A523}, 1 (1991).

\bibitem{30}
P.M. Endt,\,
Nucl.Phys. {\bf A521}, 1 (1990).

\bibitem{31}
H. Grawe, R. Shubart, K.H. Maier, D. Seweryniak,
Physica Scripta T{\bf 56}, 71 (1995).

\bibitem{32}
A. Bohr, B.R. Mottelson, {\em Nuclear Structure}, vol.1, New York
-Amsterdam, 1969.

\bibitem{Wal74}
J.D. Walecka,\, Annals of Physics {\bf 83}, 491 (1974).

\bibitem{Brock78}
R. Brockmann,\, Phys.Rev C {\bf 18}, 1510 (1978).

\bibitem{Nob79}
J.V. Noble,\, Nucl.Phys. {\bf A329}, 354 (1979).

\bibitem{Bir82}
B. L. Birbrair, L. N. Savushkin, V. N. Fomenko,
Sov.J.Nucl.Phys. {\bf 35}, 664 (1982).

\bibitem{Rein86}
P.-G. Reinhardt, M. Rufa, J. Martin, W. Greiner, J. Freidrich,\,
Z.Phys. {\bf A323}, 13 (1986).

\bibitem{Lee86}
Suk-Joon Lee, J. Fink, A.B. Balantekin, M.R. Strayer, A.S. Umar,
P.-G. Reinhard, J.A. Maruhn, W. Greiner,
Phys.Rev.Lett {\bf 57}, 2916 (1986).

\bibitem{Koepf91}
W. Koepf, P. Ring,\, Z.Phys. {\bf A339}, 81 (1991).

\bibitem{Bir98}
B.L. Birbrair, Preprint PNPI, Gatchina,  No 2234 (1998).

\bibitem{Yosh99}
S. Yoshida, H. Sagawa,\,
Nucl.Phys. {\bf A658}, 3 (1999).

\bibitem{Mac87}
R. Macleidt, K. Holinde, Ch. Elster,\,
Physics Reports {\bf 149} No1, 1 (1987).

\bibitem{9}
J. P. Omtvedt et al.,\, Nucl. Phys. A (submitted).

\bibitem{10}
K. I. Erokhina, V. I. Isakov, B. Fogelberg and H. Mach,
Preprint PNPI, Gatchina, No 2225 (1998).

\bibitem{11}
K. I. Erokhina, V. I. Isakov, B. Fogelberg and H. Mach,
Preprint PNPI, Gatchina, No 2136 (1996).

\bibitem{12}
K. I. Erokhina, V. I. Isakov, Physics of Atomic Nuclei
{\bf 57}, 198 (1994).

\bibitem{13}
K. I. Erokhina, V. I. Isakov, Physics of Atomic Nuclei
{\bf 59}, 589 (1996).


\bibitem{Nel67}
J.A. Nelder, R. Mead,\, Computer J. {\bf7}, 308 (1967).


\bibitem{37}
G.A. Leander, J. Dudek, W. Nazarewicz, J.R. Nix, Ph. Quenteen,
Phys.Rev.C {\bf 30}, 416 (1984).

\bibitem{Koura20}
H. Koura, M. Yamada, Nucl.Phys. {\bf A671}, 96 (2000).

\bibitem{Sharma95}
M.M. Sharma, G. Lalazissis, J. K\"{o}nig, P. Ring,
Phys.Rev.Lett. {\bf 74}, 3744 (1995).

\bibitem{44}
A.-M. Oros et al., in {\em Proceedings of the 6-th International Spring
Seminar on Nuclear Physics: Highlights of Modern Nuclear Structure},
S.Agata, 1998.

\end{thebibliography}
\end{document}